\documentclass[aps,prb,reprint,groupedaddress,showpacs]{revtex4-1}
\usepackage{graphicx}
\usepackage{mathrsfs} 
\usepackage{amsmath,amssymb}
\usepackage{bm}
\def\Tr{\mathop{{\rm Tr}}}
\def\i{{\bf i}}
\def\bra#1{\langle #1|}
\def\ket#1{|#1\rangle }
\def\braket#1#2{\langle #1|#2\rangle}
\def\vec#1{{\bm{#1}}}
\newcommand{\1}{\mbox{1}\hspace{-0.25em}\mbox{l}} %

\begin{document}

\title{Incommensurate Matrix Product State for Quantum Spin Systems}

\author{Hiroshi \textsc{Ueda}$^1$ and Isao \textsc{Maruyama}$^2$}
\email[]{h_ueda@riken.jp}
\email[]{maru@mp.es.osaka-u.ac.jp}
\affiliation{
$^1$Condensed Matter Theory Laboratory, RIKEN, Wako, Saitama 351-0198, Japan \\
$^2$Graduate School of Engineering Science, Osaka University, 
Toyonaka, Osaka 560-8531, Japan
}

\date{\today}

\begin{abstract}
We introduce a matrix product state (MPS) with an incommensurate periodicity by applying the spin-rotation operator of each site to a uniform MPS in the thermodynamic limit. 
The spin rotations decrease the variational energy with accompanying translational symmetry breaking and the rotational symmetry breaking in the spin space even if the Hamiltonian has the both symmetries. 
The optimized pitch of rotational operator reflects the commensurate/incommensurate  properties of spin-spin correlation functions in the $S=1/2$ Heisenberg chain and the $S=1/2$ ferro-antiferro zigzag chain. 
\end{abstract}


\maketitle

\section{Introduction}
An analysis of low-dimensional frustrated quantum spin systems beyond the mean-field approximation (MFA) is one of attractive topics in quantum mechanics, because rich quantum phases can appear due to the coexistence of frustration and strong quantum fluctuation. 
A typical example is the spin $S=1/2$ ferro-antiferro (F-AF) zigzag Heisenberg/XXZ spin chain as a theoretical model of quasi-one dimensional edge-sharing cuprates.
In theoretical studies on this quantum Hamiltonian~\cite{PhysRevB.78.144404, Sato:MPLB25, PhysRevB.80.140402, PhysRevB.74.020403, PhysRevB.76.060407, PhysRevB.76.174420, JPSJ.77.114004, PhysRevB.77.094404}, 
the exact diagonalization method (ED), the density matrix renormalization group method (DMRG)~\cite{White:PRL69-PRB48, Peschel:Springer, RevModPhys.77.259} and the infinite time-evolving block decimation method (iTEBD)~\cite{PhysRevLett.98.070201} were used as powerful methods in order to determine novel quantum phases. 

In the DMRG and the iTEBD methods, variational states take the form of a matrix product state (MPS)~\cite{Ostlund:PRL75, Rommer:PRB55} and an infinite MPS (iMPS)~\cite{PhysRevLett.98.070201}, respectively. 
When the dimension of matrices constructing the MPS is one, the MPS corresponds to the MFA.
As the dimension $m$ increases, 
the optimum variational state approaches the exact one systematically.
In addition, the MPS can handle infinite system-size directly if we suppose the spatial homogeneity of the MPS as in the iTEBD.
As a merit of the spatially uniform MPS or iMPS, there are no boundary effects which always appear in the DMRG.

In the zigzag chain~\cite{PhysRevB.78.144404, Sato:MPLB25, PhysRevB.80.140402, PhysRevB.74.020403, PhysRevB.76.060407, PhysRevB.76.174420, JPSJ.77.114004, PhysRevB.77.094404}, the helical magnetic order with incommensurate period is known to be a solution of the classical vector spin Heisenberg model which is valid in the large spin limit ($S\gg 1$). 
The incommensurate properties appear due to the geometrical frustration.  
To deal with quantum fluctuation, one can use the MPS. 
However, the spatially uniform MPS with finite dimension cannot express the helical magnetic order, because its local magnetic moment becomes spatially uniform. 
On the other hand, the DMRG can deal with a spatially inhomogeneous magnetic order, but the boundary affects incommensurate period of the order.

In this study, we propose a simple incommensurate (IC) MPS with incommensurate periodicity applying spin rotation operators\cite{Lieb1961407} to the spatially uniform MPS. 
This IC-MPS is understood naturally as a quantum generalization of the classical vector spin analysis. 
This framework is independent of the type of numerical optimization process, and it is applicable for various variational methods based on finite dimensional MPSs: DMRG~\cite{White:PRL69-PRB48, Peschel:Springer, RevModPhys.77.259}, 
the wave function predictions based on the product wave function renormalization group (PWFRG) method~\cite{JPSJ.64.4084, JPSJ.75.014003, JPSJ.77.114002, 0804.2509, JPSJ.79.044001}, 
the tensor product state (TPS)~\cite{Niggemann:ZPB104-EPJB13, Delgado:PRB64}, 
the projected entangled pair state (PEPS)~\cite{Verstraete:PRL96}, 
iTEBD~\cite{PhysRevLett.98.070201}, 
the infinite PEPS (iPEPS)~\cite{Jordan:PRL101}, 
the tree tensor network (TTN) state~\cite{Shi:PRA74}, 
the multiscale entanglement renormalization ansatz (MERA) state~\cite{Vidal:PRL101}, and so on. 
To demonstrate our light-weight modification for the uniform MPS with small dimension of matrices $m$, 
the modified Powell method~\cite{Numerical_Recipes:book} is used as a general purpose optimization method in this paper.

A pitch angle which determines an incommensurate period is a variational parameter in our approach. 
The pitch angle plays an important role in the optimization of the variational energy. 
This is caused by the finite $m$ effect, because any state can be expressed by the MPS with infinite $m$.
However, in the analysis of quantum effect starting from the classical vector spin model our approach shows a fast convergence with respect to $m$ and a result obtained by tiny $m$ is consistent with IC spin-spin correlation properties~\cite{PhysRevB.78.144404, PhysRevB.80.140402}.

The spatial periodicity and translational symmetry are recent hot topics for the MPS and its generalization~\cite{Chen:PRB82, PhysRevB.83.125104, 1103.2286, 1103.2735, JPSJ.80.023001}. 
Our previous study~\cite{JPSJ.80.023001} shown that in the spatially uniform MPS the translational symmetry breaking appeared in principal eigenvalues of its transfer matrix; that is,
the degeneracy of eigenvalues was consistent with the ground state periodicity.
This means that we need large dimension of matrices for the spatially uniform MPS with one-site periodicity
to express a magnetic ordered state with $p$-site commensurate periodicity. 
To reduce computational memory without loosing the numerical accuracy, 
$p$-site periodic MPS was effective~\cite{JPSJ.80.023001}.
However, as shown in this study, we succeed in reducing more computational memory using the IC-MPS.

This paper is organized as follows. 
In \S \ref{MPS},
we review the interaction-round-a-face (IRF)/vertex-type MPS~\cite{Baxter:book, Sierra1997505, JPSJ.80.023001} and propose the IC-MPS. 
The MFA limit of the IC-MPS is discussed in \S \ref{MFA}, where we show that the optimum vertex-type IC-MPS with $m=1$ in the $S=1/2$ Heisenberg chain is equivalent to the state from the MFA. 
In \S \ref{result}, observing the $m$ dependence of local magnetization in the $S=1/2$ Heisenberg chain, we confirm the IC-MPS takes into account the quantum fluctuation gradually by increasing $m$. 
In the same section, the effectiveness of the IC-MPS is demonstrated in the magnetization curve of $S=1/2$ Heisenberg chain and the $S=1/2$ F-AF zigzag chain under uniform magnetic field. 
Then, we discuss the reduced computational cost by applying the spin rotation in the $S=1/2$ Heisenberg chain and the C-IC change with respect to the spin-spin correlation in the zigzag chain~\cite{PhysRevB.78.144404, Sato:MPLB25, PhysRevB.80.140402, JPSJ.77.114004, PhysRevB.77.094404}. 

\section{Matrix Product state with An Incommensurate Period}\label{MPS}
Let us recall the IRF/vertex-type MPS~\cite{Baxter:book, Sierra1997505, JPSJ.80.023001}. 
An IRF-type MPS with $N$ site is
\begin{equation}
\ket{\Psi} = \sum_{\sigma} \Tr \left[ A^{ \sigma_N^{~} \sigma_1^{~} }_{ 0 } \prod_{i=1}^{N-1} A^{ \sigma_i^{~} \sigma_{i+1}^{~} }_{i} \right] \ket{\sigma},
\end{equation}
where $\sigma_i^{~}$ means the index of spin at $i$th site and $\sigma = \sigma_1^{~} \cdots \sigma_{N}^{~}$. 
The variables $A^{ \sigma_N^{~} \sigma_1^{~} }_{ 0 }$ and $A^{ \sigma_i^{~} \sigma_{i+1}^{~} }_{i}$ are $m \times m$ square complex matrices. 
The matrix $A^{ \sigma_N^{~} \sigma_1^{~} }_{ 0 }$ is called the boundary matrix~\cite{Ostlund:PRL75, Rommer:PRB55,JPSJ.79.073002,JPSJ.80.023001}. 
A vertex-type MPS is represented under the constraints: $A^{ \sigma_N^{~} \sigma_1^{~} }_{ 0 } = A^{ \sigma_N^{~} }_{ ~} A^{~}_{0}$ and 
$A^{ \sigma_i^{~} \sigma_{i+1}^{~} }_{i} = A^{ \sigma_i^{~} }_{i}$. 
To handle the thermodynamic limit ($N\rightarrow \infty$), hereafter, we treat a uniform MPS, namely $A^{ \sigma_i^{~} \sigma_{i+1}^{~} }_{i} = A^{ \sigma_i^{~} \sigma_{i+1}^{~} }_{~}$. 
As in the previous study\cite{JPSJ.80.023001}, one can treat the $p$-site periodic MPS. 

To construct an IC-MPS, we use a spin-rotational operator at each $i$th site\cite{Lieb1961407}:
\begin{equation}
\hat{R}_i(\vec{n}_i, Q_i) = \exp(- \i Q_i \hat{\vec{s}}_{i}^{~} \cdot\vec{n}_i), 
\end{equation} 
where $\i$ means a unit of pure imaginary number and $\hat{\vec{s}}_{i}^{~}$ represents the local spin operator. 
The unit vector of rotational axis and angle at each site are represented by $\vec{n}_i$ and $Q_i$, respectively. 
In this paper, we limit ourselves to the simple case of $\vec{n}_i = \vec{n}$ and $Q_i = i Q$. 

\begin{figure}[Htb]
  \centering
  \resizebox{8.5cm}{!}{\includegraphics{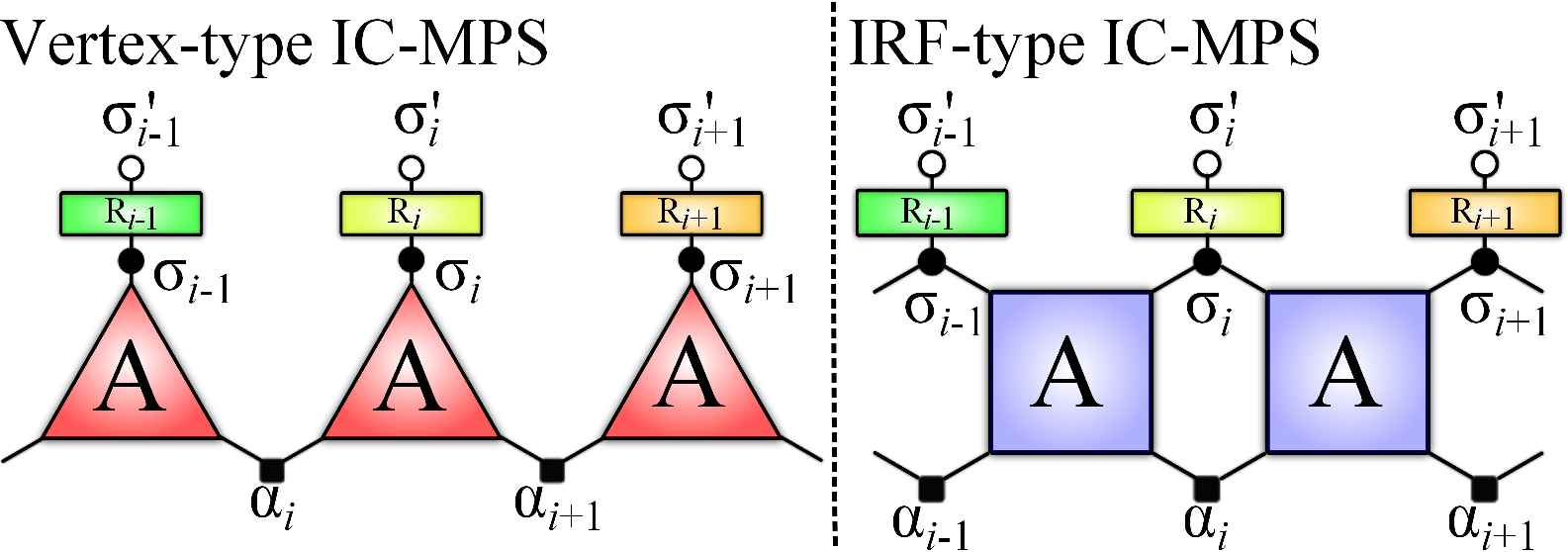}}
  \caption{(Color online) Graphical representations of vertex-type IC-MPS and IRF-type IC-MPS. 
Filled circles and filled small squares mean contraction with respect to the local spin bases $\sigma_i$ 
and the local artificial bases $\alpha_i$ of the matrix $A$, respectively.}
  \label{fig:MPS}
\end{figure}

Then, the IC-MPS is given by 
\begin{equation}
\ket{\Psi,\vec{n},Q} = \left[\prod_{i}^{} \hat{R}_i(\vec{n}, iQ) \right] \ket{\Psi} = \hat{R}_{\rm tot}(\vec{n}, Q)\ket{\Psi}. 
\end{equation}
A schematic picture of the wave function of IC-MPS is depicted in Fig.~\ref{fig:MPS}, where $R_i$ means a matrix representation of the operator $\hat{R}_i$. 

The variational energy for a Hamiltonian $\hat{H}$ is given by 
\begin{math}
e(\Psi, \vec{n}, Q)=\lim_{N\rightarrow\infty} E(\Psi, \vec{n}, Q)/N
\end{math}
with
\begin{eqnarray}
E(\Psi, \vec{n}, Q)&=&\bra{\Psi, \vec{n}, Q} \hat{H} \ket{\Psi, \vec{n}, Q} / \braket{\Psi, \vec{n}, Q}{\Psi, \vec{n}, Q}
\nonumber
\\
&=&\bra{\Psi} \hat{H}(\vec{n}, Q) \ket{\Psi} / \braket{\Psi}{\Psi}  
,
\\
\hat{H}(\vec{n}, Q) &=& 
\hat{R}^\dagger_{\rm tot}(\vec{n}, Q) \hat{H} \hat{R}_{\rm tot}(\vec{n}, Q)
,
\end{eqnarray}
where $\hat{H}(\vec{n}, Q)$ is the spin-rotated Hamiltonian.  Then, hereafter, we just consider $\hat{H}(\vec{n}, Q)$.

For general $\vec{n}_i$ and $Q_i$, the important characters of the spin-rotated operator are summarized below. 
The rotated local spin operator in general spin $S$ is given by 
\begin{equation}
\hat{ \vec{s} }_i (\vec{n}_i, Q_i) = \hat{R}^\dagger_{i} (\vec{n}_i, Q_i) \hat{ \vec{s} }_i \hat{R}_i(\vec{n}_i, Q_i) = {\bf D}(\vec{n}_i, Q_i) \hat{ \vec{s} }_i, 
\end{equation}
where the three dimensional matrix ${\bf D}(\vec{n}_i, Q_i)$ is given by
\begin{eqnarray}
\big[ {\bf D} (\vec{v}, q) \big]^{~}_{\eta \eta'_{~}} & = & 
v^{~}_{\eta} v^{~}_{\eta'_{~}}  + (\delta^{~}_{\eta \eta'_{~}} - v^{~}_{\eta} v^{~}_{\eta'_{~}}) \cos q \nonumber \\
& & -  \sin q \sum_{\eta^{\prime \prime}_{~}} \epsilon_{\eta \eta'_{~} \eta^{\prime \prime}_{~}} v^{~}_{\eta^{\prime \prime}_{~}}
, 
\label{D} 
\end{eqnarray}
for the unit vector $\vec{v}$.
Symbols $\delta^{~}_{\eta \eta'_{~}}$ and $\epsilon_{\eta \eta'_{~} \eta^{\prime \prime}_{~}}$ represent the Kronecker delta and the Levi-Civita symbol, respectively, where $\eta = x,y,z$.  
From Eq. (\ref{D}), we can immediately obtain the relation ${\bf D}(\vec{n}_i, Q_i)^{\rm t} = {\bf D}(\vec{n}_i, -Q_i)$. 

For the simple case of $\vec{n}_i = \vec{n}$ and $Q_i = i Q$,
one can prove the following equation: 
\begin{eqnarray}
\hat{ \vec{s} }_i (\vec{n}, iQ) \cdot \hat{ \vec{s} }_{i+\ell} (\vec{n}, (i+\ell)Q)  
= \hat{ \vec{s} }_{i} \cdot \hat{ \vec{s} }_{i+\ell}(\vec{n},\ell Q). 
\end{eqnarray}
The vanishing of position dependence simplifies the calculation of the Heisenberg Hamiltonian.
For the $S=1/2$ Heisenberg chain defined by
\begin{equation}
\hat{H}_1 
= \sum_{i} ^{~} \hat{ \vec{s} }_i^{~} \cdot  \hat{ \vec{s} }_{i+1}^{~}
,
\end{equation}
the spin-rotated Hamiltonian $\hat{H}_1(\vec{n}, Q)$ is written as
\begin{eqnarray}
\hat{H}_1(\vec{n}, Q) 
= \sum_i \hat{\vec{s}}_i \cdot \hat{ \vec{s} }_{i+1}(\vec{n}, Q) 
= \sum_i \hat{h}_i(\vec{n}, Q) 
.
\label{ham_s12}
\end{eqnarray} 
This Hamiltonian has the translational symmetry.  Then, we apply the same uniform MPS used in the previous study~\cite{JPSJ.80.023001}.
If the artificial translational-symmetry breaking does not occur, we can neglect the boundary matrix and
the local energy $e_i=\bra{\Psi} \hat{h}_i(\vec{n}, Q) \ket{\Psi}$ becomes independent of position $i$ in the thermodynamic limit.
The translational symmetry of the spin-rotated Hamiltonian is recovered even for the zigzag and bilinear-biquadratic Heisenberg chain for general spin $S$.

It should be noted that we can deal with the case that the spin-rotated Hamiltonian does not have the translational symmetry.
In this case, we can calculate the variational energy by using the translational symmetry of the MPS $\ket{\Psi}$, 
because the position dependence of the local energy $e_i$ can be expanded as 
\begin{equation}
e_i=e^{(0)} + \sum_{k\neq 0} e^{(k)} \exp(\i i k Q)
\label{expand_energy}
\end{equation}
and only $e^{(0)}$ gives non-zero contribution after taking the summation $\sum_i e_i$ {\it if $Q$ is not commensurate}.
For commensurate $Q$, we must consider the contribution from $e^{(k)}$ for $k\neq 0$.
Of course, one can treat more general position-dependent rotations, for example $Q_{2i} = 2iQ_{\rm a}$ and $Q_{2i+1} = (2i+1)Q_{\rm b}$, 
where the expansion as in Eq. (\ref{expand_energy}) becomes more complex, namely 
$e_i=e^{(0,0)} + \sum_{(k_{\rm a},k_{\rm b}) \neq (0,0)} e^{(k_{\rm a},k_{\rm b})} \exp(\i i (k_{\rm a} Q_{\rm a} + k_{\rm b} Q_{\rm b}))$. 

\section{Mean Field Approximation Limit} \label{MFA}
We derive the mean-field limit of this method, which is realized by the vertex-type IC-MPS with $m=1$.
In this limit, we can neglect the boundary $A_0$ which has only trivial two roles: normalization and phase factor. 
Then, the MPS becomes a direct product state
\begin{math}
\ket{\Psi} = \sum_{\sigma} \prod_i(A^{\sigma_i} \ket{\sigma_i}) 
,
\end{math}
expressed by two complex variables, $A^{\uparrow}$ and $A^{\downarrow}$ in $S=1/2$ systems.
As a normalization, we assume $\sum_{\sigma_i} |A^{\sigma_i}|^2 = 1$.

To show that the mean-field limit corresponds to the classical vector spin model, we consider the Heisenberg Hamiltonian, $H_1$. 
The variational energy is given by 
\begin{equation}
e(\Psi,\vec{n}, Q)
=  \vec{M}\cdot ({\bf D}(\vec{n}, Q) \vec{M})
,
\end{equation}
with an expectation value of local magnetic moment
\begin{math}
\vec{M} = \sum_{\sigma,\sigma'} A^{\sigma *} A^{\sigma'} \bra{\sigma}  \hat{\vec{s}} \ket{\sigma'} 
.
\end{math}
The local magnetization is obtained by $|\vec{M}| = \sqrt{ \sum_\alpha \bra{\Psi} \hat{s}^\alpha \ket{\Psi}^2}$. 
After the optimization for fixed $Q$,
one can obtain
\begin{equation}
e(Q)=\min_{\Psi,\vec{n}} e(\Psi,\vec{n}, Q) =\cos{Q}/4
.
\end{equation}
Then, the optimization of $e(Q)$ gives the N\'eel-type solution $Q=\pi$.
We stress again that 
this energy gain of $e(Q)-e(0)$ is due to finite $m$,
because any state can be expressed by the uniform ($p=1, Q=0$) MPS accurately if we have enough large dimension $m$ for the MPS.
This finite-dimensionality also causes $|\vec{M}|=1/2$ which is always proved for any state in the mean-field limit,
while it is known that the exact ground state does not have the magnetization at zero magnetic field.
In this sense, the mean-field limit corresponds to the classical vector spin model.
In fact, as shown in \S \ref{result}, when we increase $m$ to express quantum fluctuation or entanglement, the local magnetization 
$|\vec{M}|$ obtained after the optimization decreases and approaches to the exact value.

\section{Numerical Result and Discussion} \label{result}
Before showing results, we summarize details of our numerical calculation.
We prepare the $m$-dimensional complex matrix $A^{\sigma_i, \sigma_{i+1}}_{}$ for the IRF-type uniform MPS. 
The rotational axis $\vec{n}$ is fixed as $(0,0,1)$ to conserve translational symmetry of the rotated uniaxial Hamiltonian with the longitudinal magnetic-field $H_z$ applied in $z$-axis. 
The pitch $Q$ and $A^{\sigma_i, \sigma_{i+1}}_{}$ are optimized so that the variational energy $e$
for a given Hamiltonian $\hat{H}$ becomes minimum by using the modified Powell method~\cite{Numerical_Recipes:book}. 
The number of optimization parameters in the IRF-type IC-MPS under fixed rotational axis is $2 d^2 m^2 + 1 $, where the coefficient $2$ comes from using complex numbers and $d$ is the degree of freedom (DOF) of local spin, namely 2 in this work. The term g+1h means the DOF of the wave number $Q$. 
In the optimization, $10 - 2000$ initial states are prepared and optimized in each Hamiltonian parameter to avoid obtaining a local minimum.

\begin{figure}[Htb]
  \centering
  \resizebox{7cm}{!}{\includegraphics{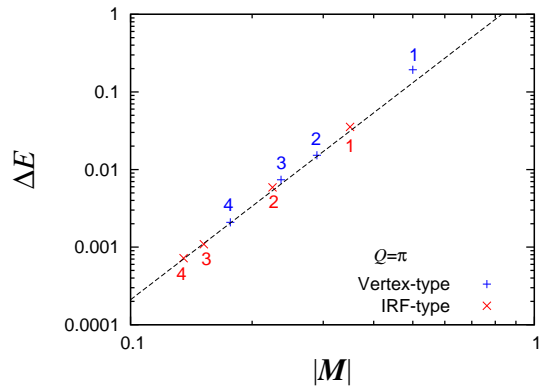}}
  \caption{(Color online) Energy error $\Delta E$ as a function of local magnetization $|\vec{M}|$ in the $S=1/2$ Heisenberg chain $\hat{H}_1$. The number nearby each symbol means the matrix dimension $m$. The broken line is guide for the eyes showing the power low decay of $\Delta E \propto |\vec{M}|^4$.} 
  \label{fig:spin_dens}
\end{figure}

The MPS gradually takes account of the quantum fluctuation of the local magnetic moment in the S=1/2 Heisenberg chain with increasing $m$ as shown in Fig. \ref{fig:spin_dens}.
The rotational angle $Q = \pi$ is obtained after the optimization. 
The energy error $\Delta E$ means the difference between optimized variational energies as function of $m$ and the exact energy $-\ln2 + 1/4$~\cite{Takahashi}. 
The energy error and the local magnetization $|\vec{M}|$ are monotonically decreasing with respect to $m$. 
We confirm that the IRF-type MPS can deal with non-zero quantum fluctuations even if $m=1$, while the vertex-type MPS with $m=1$ gives the mean-field result.
This is an advantage of using IRF-type MPS. 

\begin{figure}[Htb]
  \centering
  \resizebox{7cm}{!}{\includegraphics{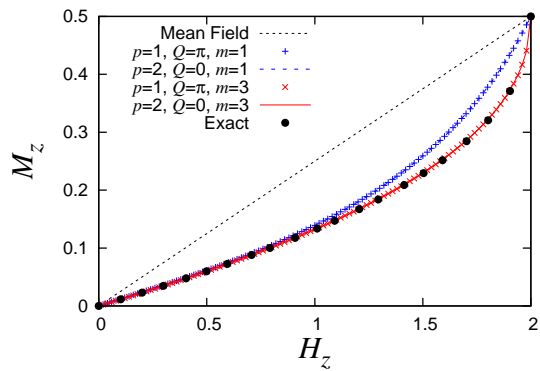}}
  \caption{(Color online) Magnetization $M_z$ curve as a function of uniform magnetic field $H_z$ in the $S=1/2$ Heisenberg chain $\hat{H}_1$. } 
  \label{fig:mag_s12}
\end{figure}

The magnetization $M_z$ in the $S=1/2$ Heisenberg chain with the magnetic-field $H_z$ is shown in Fig. \ref{fig:mag_s12}. 
As reference data, we show the exact result for $S=1/2$ from the Bethe ansatz~\cite{Takahashi} and 
the result from two-site modulated MPS, named $p=2$, $Q=0$, in our previous study~\cite{JPSJ.80.023001}. 
While the mean field result fails to obtain the correct criticality near the fully saturated point,
results for $m=3$, which is not so large dimension, show enough accuracy.

This increasing of $m$ leads to a great improvement in the accuracy of estimating the magnetization curve. 
The relative error of the magnetization curve from the IC-MPS with $m=3$ in Fig. \ref{fig:mag_s12} is smaller than 3\% even though the error of local magnetization $|\vec{M}|$ is of the order of that from the MFA as shown in Fig. \ref{fig:spin_dens}; that is, the absolute error of $M_z$  in Fig. \ref{fig:mag_s12} is less than $0.001$ even though the error of $|\vec{M}|$ in Fig. \ref{fig:spin_dens} is about 0.1.

Moreover, in both cases of $m=1$ and $m=3$, the data from IC-MPS with $Q=\pi$ agree with that of the $p=2, Q=0$ MPS. 
This means that the number of optimization parameter is reduced by 50\% compared to the previous study.

\begin{figure}[Htb]
  \centering
  \resizebox{6.6cm}{!}{\includegraphics{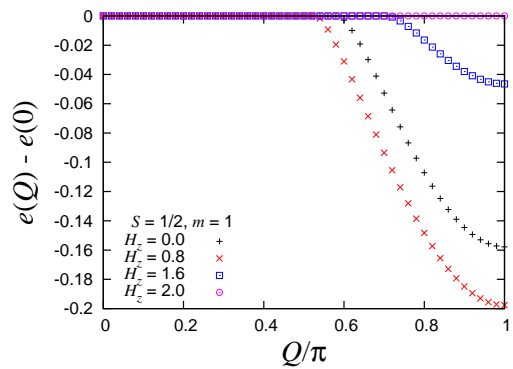}}
  \caption{(Color online)  Rotational parameter $Q$ dependence of the variational energy in the $S=1/2$ Heisenberg chain $\hat{H}_1$. }
  \label{fig:optimum_rotational_pitch}
\end{figure}

How the rotational pitch $Q=\pi$ is stabilized by the energy gain, $e(Q)-e(0)$, is shown in Fig.~\ref{fig:optimum_rotational_pitch}. 
This figure clearly shows that the variational energy becomes minimum at $Q=\pi$ for any magnetic field except for $H_z=2.0$ in the perfect-ferro region.
Compared with 
\begin{math}
e(Q)=\cos{Q}/4  
\end{math}
for $H_z=0$ in the mean-field limit,
there is the flat energy region in small $Q$ region for $H_z=0$. 
In the flat region, we confirm the state is a superposition of the N\'eel state, namely the linear combination of $\ket{\uparrow\downarrow\uparrow\ldots}$ and $\ket{\downarrow\uparrow\downarrow\ldots}$~\cite{JPSJ.80.023001}.
This state is invariant with respect to the spin rotation along z-axis trivially. 
The origin of the flat region is the quantum fluctuation of the N\'eel state.
This quantum fluctuation can be expressed by the IRF even in $m=1$.

Finally, we discuss the periodicity change appearing in the $S=1/2$ F-AF ($J_1<0$ and $J_2 >0$) zigzag Heisenberg chain with uniform longitudinal magnetic field,
\begin{equation}
\hat{H}_2 = \sum_{i} \left( \sum_{k=1,2} J_k \hat{\vec{s}}_i \cdot \hat{\vec{s}}_{i+k} - H_z \hat{s}^z_{i} \right)
. 
\end{equation}
The longitudinal magnetic field $H_z$ is taken as $0$ and $0.1$ in this analysis. 
At $H_z=0$, there is a C-IC change at $J_1/J_2=-4$. 
The commensurate state for $J_1/J_2<-4$ is a ferromagnetic state while characterization of the ground state for $J_1/J_2>-4$ is a difficult task.
Recent study~\cite{Sato:MPLB25} pointed out that the ground state for $J_1/J_2>-4$ is the Haldane-dimer phase, 
which is characterized by a generalized string order parameter, where ordinal spin-spin correlations behave incommensurately~\cite{PhysRevB.80.140402}. 
This incommensurate behavior is also found in the VC phase for non-zero magnetic fields~\cite{PhysRevB.78.144404}.

\begin{figure}[Htb]
  \centering
  \resizebox{7cm}{!}{\includegraphics{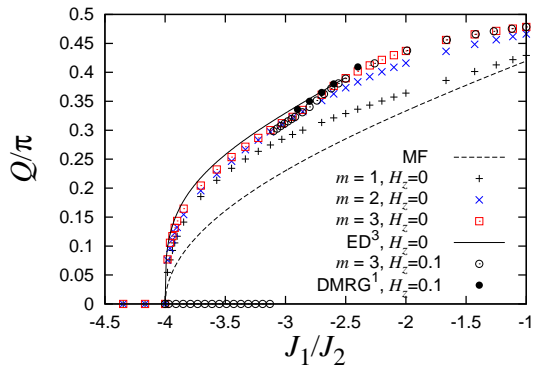}}
  \caption{(Color online) Rotational parameter $Q$ as a function of $J_1/J_2$ in the zigzag chain $\hat{H}_2$. }
  \label{fig:incomme}
\end{figure}

To demonstrate our approach for the C-IC change, optimized pitch $Q$ is calculated for this frustrated Hamiltonian $\hat{H}_2$ as shown in Fig. \ref{fig:incomme}. 
In this figure there are three kinds of the reference data.  First, the broken line is the result of the mean-field approximation, $Q={\rm arccos}(-J_1/4J_2)$.    Second, the solid line is the fitting line for the location of the maximum of the zero field spin structure factor with the ED~\cite{PhysRevB.80.140402}, where $Q\propto (J2-1/4)^{0.29}$.  Finally, the filled circles are the result of  the DMRG at $M_z=0.05$ in the VC phase~\cite{PhysRevB.78.144404}.
For $H_z=0$, the pitch $Q$ approaches $\pi/2$ with increasing $J_1/J_2$ more rapidly than that of the mean field approximation due to taking into account the quantum fluctuation by growing $m$. 
On the other hand, the C-IC change point is completely converged at $J_1/J_2 = -4$. 
We find the pitch in $m=3$ is comparable with the result from ED~\cite{PhysRevB.80.140402} in $J1/J2 \leq -2.5$. 
Around the transition point, the pitch is well converged with respect to $m$ in this scale. 
The $m$ dependence becomes gradually large with increasing $J_1/J_2$, where the frustration due to $J_2$ becomes also gradually large. 

For $H_z=0.1$, the pitch $Q$ depicted by the open circle in Fig. \ref{fig:incomme} has a jump around $J_1/J_2 = -3.1$,
which is very close the SDW$_3$--VC phase transition point~\cite{PhysRevB.78.144404}. 
Unfortunately, the pitch $Q$ for small $m$ fails to capture the SDW$_3$ and SDW$_2$ states, where there are the Ferro--SDW$_3$ phase transition at $J_1/J_2 \sim -3.3$ and the VC--SDW$_2$ phase transition at $J_1/J_2 \sim -2$~\cite{PhysRevB.78.144404}.
In the same meaning, the characterization of the ground state at $H_z=0$ is difficult for our method at this stage.
Nevertheless, a notable point is that the incommensurate pitch $Q$ of the IC-MPS for the VC phase in $J_1/J_2 > -3.1$ shows reasonable agreement with the DMRG result.
For $J_1/J_2 > -3.1$, $Q$ of the IC-MPS is nearly independent of $H_z$, which is also consistent with the DMRG analysis~\cite{PhysRevB.78.144404}.

\section{Summary}
In summary, we introduced the IRF-type MPS with the incommensurate pitch parameter $Q$ and the rotational axis $\vec{n}$ as a generalization of the uniform MPS,
which can be used for various variational methods based on the MPS.
Two parameters $Q$ and $\vec{n}$ allow us to evaluate an incommensurability of the spin chain in the thermodynamic limit directly.
Our approach with small dimension of matrices is connected to the classical vector spin Heisenberg model which is valid in the large spin limit ($S\gg 1$). 
For the exact ground state,
the helical magnetic order obtained in the classical limit is expected to be destroyed by quantum fluctuations in the quantum limit $S=1/2$. 
However, we emphasize quantum effects on some quantities are rapidly converged with respect to the matrix dimension.
Our approach opens a way to a light-weight analysis based on the classical vector spin model to include quantum fluctuation. 
Using this approach, one can treat translational symmetry broken states, such as the helical magnetic order, in the thermodynamic limit, 
which cannot be handled by known iMPS with translational symmetry.

We demonstrated the efficiency of this IRF-type IC-MPS in two types of Hamiltonians:
i) the magnetization in the $S=1/2$ antiferro-magnetic Heisenberg chain under uniform magnetic field, 
and ii) the C-IC change in $S=1/2$ F-AF Heisenberg zigzag chain under uniform magnetic field. 
In the former Hamiltonian, we have succeeded in obtaining the same result as two-site modulated MPS.
This means 50\% reduction of the number of optimization parameters. 
In the latter Hamiltonian, we have succeeded in detection of the C-IC change of correlation properties with increasing $m$. 
The pitch $Q$ near the C-IC transition point is immediately converged with respect to $m$
and shows reasonable agreement with the ED study~\cite{PhysRevB.80.140402} and the DMRG study~\cite{PhysRevB.78.144404}, despite small $m$.

On the other hand, the sufficiently converged $Q$ is not obtained around the strongly frustrated region, namely $|J_1| \sim J_2$.  
To discuss the details of $Q$, analysis with larger $m$ are necessary.  
For this problem, we can apply other optimization methods using the Trotter decomposition~\cite{PhysRevLett.98.070201}, the matrix product operator (MPO)~\cite{1367-2630-12-2-025012}, and the time-dependent variational principle (TDVP)~\cite{PhysRevLett.107.070601} to updating the MPS under given $Q_i$ and $\vec{n}_i$. 
We stress again that the framework of IC-MPS is independent of the type of numerical optimization process. 
In this paper, the modified Powell method was chosen as an optimization method because it is a general purpose method and all parameters are optimized easily.
The modified Powell method is enough to clarify to effectiveness of our light-weight modification but becomes a bottleneck when we increase $m$. 
To study larger $m$, convergence properties and numerical efficiencies of these updating methods should be discussed. 
This is one of future problems.

Another future issue is to change the constraint of the rotational axis and pitch parameter in order to represent the magnetization plateau state or the SDW state, for example $\vec{n}_i=\vec{n}_{{\rm mod}[i, p]}$ and $Q_i=Q_{{\rm mod}[i, p]}$. 
As another application, we have already performed other C-IC correlation properties change in the bilinear-biquadratic spin $S=1$ chain~\cite{1111.3488}, and succeeded to detect the C-IC change with the IC-MPS, which cannot be detected by the mean filed approximation. 

This method uses the spin-rotational operator which maps the classical helical state to the perfect-ferro state. 
The uniform direct product state including the perfect-ferro state can be always described by the uniform MPS with $m=1$. 
A generalization of this method is to find another kind of spin-rotational operator which maps the ground state to the uniform direct product state.
The role of this operation is similar to disentanglers in the MERA~\cite{Vidal:PRL101}. 
In this sense, it is interesting to consider the valence bond solid (VBS) state which cannot be rotated by the spin-rotational operators\cite{Lieb1961407}
and is known to have the Kennedy-Tasaki (KT) transformation which converts the string order to the ferromagnetic order 
as a global topological disentangler~\cite{PhysRevB.83.104411,1109.4202}.

In general, the classical magnetic order can be appeared easily in higher-dimensional systems. 
In this case, the spin rotation becomes effective. 
Moreover, in higher-dimensional systems, the dimension of the matrix/tensor is restricted due to the computational resources.
Then, a small $m$ analysis based on our approach is an interesting approach for the incommensurate TPS for 2D quantum spin systems, which is one of future issues. 
Not only for the TPS, our approach can be applied to various methods.

\begin{acknowledgments}
We acknowledge discussions with S.~Miyahara.
This work was supported in part by a Grant-in-Aid for JSPS Fellows and Grant-in-Aid No. 20740214, Global COE Program (Core Research and Engineering of Advanced Materials-Interdisciplinary Education Center for Materials Science) from the Ministry of Education, Culture, Sports, Science and Technology of Japan. 
\end{acknowledgments}

\end{document}